\documentclass{aastex}          %% The manuscript based on AASTeX v5.x
\usepackage{spr-astr-addons}    %% mimicing ASTR journal style
\begin{document}
%% Article title
%
\title{A Young Multipolar Planetary Nebula in the Making -- \\ 
          IRAS 21282+5050}

%% Running heads
\shorttitle{Study of Planetary Nebula IRAS 21282+5050}
\shortauthors{<Hsia et al.>}

%% Author and Affilations
\author{Chih-Hao Hsia\altaffilmark{1, 3}}
\author{Yong Zhang\altaffilmark{2,3}}
\author{Sun Kwok\altaffilmark{3,4}}\email{skwok@eoas.ubc.ca} %% non-output
%\and 
\author{Wayne Chau\altaffilmark{3,5}}
%\affil{}

%% Alternate Affilations
\altaffiltext{1}{State Key Laboratory of Lunar and Planetary Sciences, Macau University of Science and Technology, Taipa, Macau 999078, China}
\altaffiltext{2}{School of Physics and Astronomy, Sun Yat-Sen University Zhuhai Campus, Tangjia, Zhuhai, 519082, China}
\altaffiltext{3}{Laboratory for Space Research,  The University of Hong Kong, Hong Kong, China}
\altaffiltext{4}{Department of Earth, Ocean, and Atmospheric Sciences, University of British Columbia, Vancouver, Canada}
\altaffiltext{5}{Department of Physics, The University of Hong Kong, Hong Kong, China}

\noindent{Email address of corresponding author Sun Kwok: skwok@eoas.ubc.ca}\\

%% Abstract
\begin{abstract}
We present high-angular-resolution {\it Hubble Space Telescope (HST)} optical and near-infrared imaging of 
the compact planetary nebula (PN) IRAS 21282+5050. 
Optical images of this object reveal several complex morphological structures including three pairs of bipolar lobes and an elliptical shell lying close to the plane of the sky. 
From near-infrared observations, we found a dust torus oriented nearly perpendicular to the major axis of elliptical shell. 
The results suggest that IRAS 21282+5050 is a multipolar PN, and these structures developed early during the post 
asymptotic-giant-branch (AGB) evolution. 
From a three-dimensional (3-D) model, we derived the physical dimensions of these apparent structures. 
When the 3-D model is viewed from different orientations, IRAS 21282+5050 shows similar apparent structures as other multipolar PNs. 
Analysis of the spectral energy distribution and optical spectroscopic observations of the nebula suggests the presence of a cool companion to the hot central star responsible for the ionization of the nebula.
Whether the binary nature of the central star has any relations with the multipolar structure of the nebula needs to be further investigated.

\end{abstract}

%% Keywords
%\keywords{}

%%  Please use labels (\label, \ref) for section, figure, table, 
%%  equation  reference. Use \cite for bibliography references.
%
%\section{}%\label{s:?}
%\subsection{}%\label{ss:?}
%\subsubsection{}%\label{sss:?}

\section{Introduction}

Planetary nebulae (PNs) are traditionally known for their simple spherical, shell-like structures. Their morphological structures are often classified as round, elliptical, or bipolar \citep{Balick87}. After the development of high-dynamic-range CCD imaging, especially after the launch of the {\it Hubble Space Telescope (HST)}, many complex internal structures have been discovered. These include multipolar lobes, circular concentric arcs, two-dimensional (2-D) rings, jets, knots, ansae, extended halos, and equatorial tori.

Multipolar nebulae are defined as objects that possess at least two pairs of axial symmetric structures. The first multipolar PN were discovered in the Instituto de Astrof\'{i}sica de Canarias  morphological survey of Galactic PNs \citep{Manchado96}.  Five objects (M2-46, K3-24, M1-75, M3-28, M4-14) were identified by \citet{Manchado96} as PNs with quadrupolar structures.
Multipolar PN found in subsequent deep and high-dynamic-range optical imaging include Hen 2-47 and M 1-37 \citep{Sahai00}, NGC 6881 \citep{Kwok05}, NGC 6072 \citep{Kwok10b}, NGC 6644 \citep{Hsia10}, NGC 5189 \citep{Sabin12}, Kn 26 \citep{Guerrero13}, NGC 6058 \citep{Guillen13}, NGC 6309 \citep{Rubio15}, Hen 3-1333 and Hen 2-113 \citep{Dan15}. Under earlier classification schemes, these objects were classified as ``elliptical'' and/or ``bipolar''. 
Even large, well-known PN, e.g., NGC 2440 \citep{Lopez98} and NGC 7026 \citep{Clark13}, are now known to be multipolar. 

Recent classification schemes find 12$\%$ of 150 post asymptotic-giant-branch (AGB) stars \citep{Manchado11} and 20$\%$ of 119 young PNs \citep{Sahai11} to be multipolar. A study of multipolar or quadrupolar PNs by \citet{Hsia14} suggests that these multipolar structures could be the result of interactions between previously ejected AGB winds and later-developed fast winds, the presences of the structures are ascribed to the multiple phases of fast winds separated by time or directional variations. What is the nature and formation of multipolar lobes? 
Although some studies have discussed this issue \citep{Garcia10, Kwok10a, Velazquez12}, the origin and exact physical mechanisms responsible for these structures are still unresolved.

\begin{table*}
	\caption{Log of HST observations for IRAS 21282+5050.}
	\centering
	\begin{tabular}{llccccc}
		\hline\hline
		Instrument & Filter & $\lambda_{c}$ (\AA~) & $\triangle$ $\lambda$ (\AA~) & Exposures (s) & Observation Date & Program ID \\
		\hline
		& & \multicolumn{4}{c}{Visible} & \\
		\hline
		ACS & F606W & 5919 & 2342 & 5$\times$1 & 2002 Dec 07 & 9463  \\
		& F606W & 5919 & 2342 & 60$\times$1 & 2002 Dec 07 & 9463  \\
		& F606W & 5919 & 2342 & 250$\times$2 & 2002 Dec 07 & 9463  \\
		WFPC2 & F656N & 6564 & 22 & 20$\times$2 & 1999 Aug 23 & 8345  \\ 
		& F656N & 6564 & 22 & 140$\times$2 & 1999 Aug 23 & 8345  \\
		& F656N & 6564 & 22 & 400$\times$2 & 1999 Aug 23 & 8345  \\
		\hline
		& & \multicolumn{4}{c}{Near-infrared} & \\
		\hline
		NICMOS & F160W & 15931 & 4030 & 32$\times$12 & 1997 Sep 21 & 7365  \\
		& F205W & 20406 & 6125 & 32$\times$12 & 1997 Sep 21 & 7365  \\
		\hline
	\end{tabular}
	\label{tab1}
\end{table*}

The PN IRAS 21282+5050 (PNG 093.9-00.1, hereafter referred to as IRAS 21282) was first discovered by {\it Infrared Astronomical Satellite (IRAS)} for its strong infrared emission (UIE) features (7.7, 8.6, 11.3, and 12 $\mu$m) in its {\it IRAS} low-resolution spectra \citep{Cohen85, Jourdain86}. 
These features are part of a family of unidentified infrared emission bands (3.3, 3.4, 6.2, 7.7, 8.6, 11.3, 12.6 $\mu$m) commonly seen in carbon-rich PNs and PPNs \citep{Kwok99}. 
In addition to common UIE bands, weaker features at 3.46 and 3.52 $\mu$m are also seen \citep{Jourdain86, Jourdain90, Nagata88}. 
High-spectral-resolution observations have identified features at 3.40, 3.46, 3.52 as due to aliphatic C--H stretch of methyl and methylene groups and the 3.45 $\mu$m as due to C--H stretch of an aldehydes group \citep{Hrivnak07}.

In this paper, we present results of our morphological study for IRAS 21282 based on optical and near-infrared images taken with {\it HST}. These images reveal the unique morphology and properties of the PN. The observations and data reduction procedures are described in $\S$ 2. In $\S$ 3 we present the results of imaging and spectroscopy in the optical and near-infrared for this source. From photometric and spectroscopic observations from radio to ultraviolet (UV), we constructed a spectral energy distribution (SED) and discuss the energetics of the central source in $\S$ 4. A comparison of our observations with a three-dimensional model are presented in $\S$ 5. A discussion of the implications based on a comparison of studied multipolar PNs is given in $\S$ 6. Finally, a conclusion is given in $\S$ 7.

\section{Observations and data reduction}\label{s2}

\subsection{{\it HST} optical imaging}

\begin{figure}
	\begin{center}
		\includegraphics[width=0.5\textwidth]{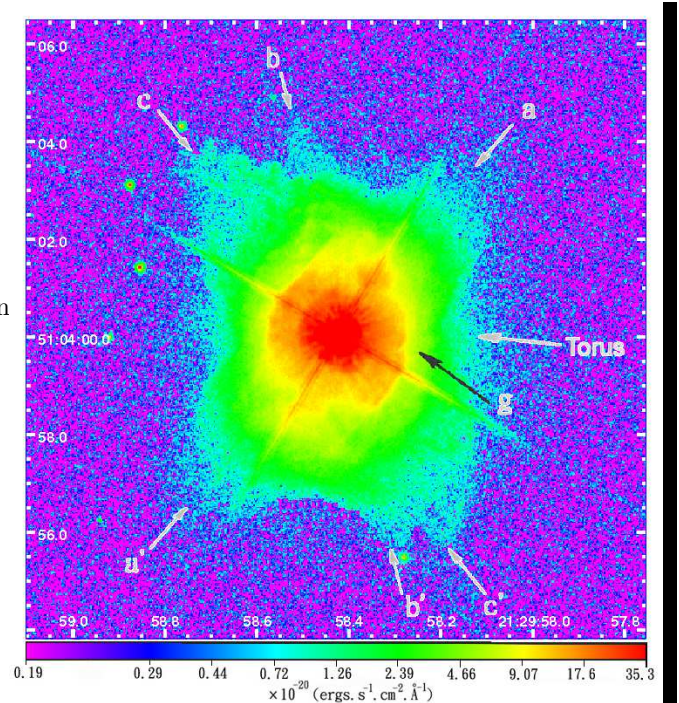}
	\end{center}
	\caption{The complex morphological structure of IRAS 21282+505 is shown in this {\it HST/ACS} F606W broad-band image. At the center of the nebula is a bright cylindrical shell (labeled  ``$g$'').  In the outer regions, three pairs of bipolar lobes (marked as $a-a^\prime$, $b-b^\prime$, and $c-c^\prime$) can be identified.
Along the E-W direction is an extended region which could be the counterpart of the torus seen in the NICOMS image (Fig.~\ref{nicmos}).
The intensity display is on a logarithmic scale and the scale bar is given at the bottom in units of ergs cm$^{-2}$ s$^{-1}$ \AA~$^{-1}$.}
\label{acs}
\end{figure}

We have analyzed the high-resolution optical images on IRAS 21282 retrieved from the Space Telescope Science Archive. 
The broad-band images were obtained under program 9463 (PI: R. Sahai) using the Advanced Camera for Surveys (ACS) on {\it HST}. The object was observed with broad F606W (V-band) filter ($\lambda_{c}$ = 5919 \AA~, $\triangle\lambda$ = 2342 \AA) on the High Resolution Channel (HRC) on 2002 December 7, which provides a 26$\arcsec$ $\times$ 29$\arcsec$ field of view (FOV) at a spatial resolution of 0.$\arcsec$027 pixel$^{-1}$. The actual observations were made with different exposure times (from 5 s to 250 s) to allow for the imaging of both the bright central region and the faint outer parts. 

In addition, we also retrieved narrow-band observations using the Wide Field Planetary Camera 2 (WFPC 2) on {\it HST} through the observation program 8345 (PI: R. Sahai) on 1999 August 23. The nebula was imaged by the Planetary Camera (PC), which has a FOV of 36.$\arcsec$8 $\times$ 36.$\arcsec$8 with a pixel scale of 0.$\arcsec$045 pixel$^{-1}$. The narrow-band imaging observations were made with narrow-band F656N (H$\alpha$) filter ($\lambda_{c}$ = 6564 \AA~, $\triangle\lambda$ = 22 \AA). 

The standard {\it HST} pipeline calibration has been applied to all data. Successful bias subtraction and flat-field correction were performed. Data were taken in two-step dithered positions to enhance spatial sampling and cosmic rays removal by using the task {\bf crrej} in the STSDAS package of IRAF. The journal of observations is summarized in Table \ref{tab1}. The processed F606W and F656N  (H$\alpha$) images with a total exposure time of 565 s and 1120 s are shown in Figures~\ref{acs} and \ref{ha}.

\begin{table*}
	\caption{Summary of optical spectral observation of IRAS 21282+5050}
	\centering
	\begin{tabular}{ccccc}
		\hline\hline
		Observation Date &  Wavelength Range &  Resolution &  Width of Slit & Integration Time \\
		&            (\AA~) & (\AA~pixel$^{-1}$) & (arcsec) & (s) \\
		\hline
		2011 Sep 26 & 6300 - 7600 & 1.0 & 3.0 & 2100 $\times$ 2 \\
		2011 Sep 26 & 6300 - 7600 & 1.0 & 3.0 & 1500  \\
		2011 Sep 26 & 6300 - 7600 & 1.0 & 3.0 & 1200  \\
		\hline
	\end{tabular}
	\label{tab2}
\end{table*}

\subsection{{\it HST} near-infrared imaging}

High-resolution near-infrared images of IRAS 21282 were obtained by the Near-Infrared Camera Multiobject Spectrometer 
(NICMOS) on {\it HST} through the observation program 7365 (PI: W. B. Latter). The data were all obtained on 1997 September 21. Two broad-band filters F160W ($\lambda_{c}$ = 15,931 \AA~, $\triangle\lambda$ = 4,030 \AA) and F205W ($\lambda_{c}$ = 20,406 \AA~, $\triangle\lambda$ = 6,125 \AA) were employed for these images. The 1.6 $\mu$m (F160W) observations were made with Camera 1 (NIC1), which has a FOV of 11$\arcsec$ $\times$ 11$\arcsec$ with a spatial resolution of 0.$\arcsec$043 pixel$^{-1}$. 
Another observation 
(F205W) was carried out with Camera 2 (NIC2), which provides a 19.$\arcsec$2 $\times$ 19.$\arcsec$2 FOV at a pixel scale of 0.$\arcsec$075 pixel$^{-1}$. The total exposure times for these two observations (F160W and F205W) are both 384 s. The data were reduced and calibrated by the standard {\it HST} NICMOS pipeline. Standard flat-field correction and bias subtraction were performed. The multiple images of each band were combined together using the {\bf calnicB} task in the STSDAS package of IRAF. A summary of NICMOS observations is also given in Table \ref{tab1}.

\subsection{Optical spectroscopy}

Optical mid-resolution spectra of the nebula were performed on the night of 2011 September 26 with the {\it 2.16 m Telescope} on the Xing-Long station of the National Astronomy Observatories of China (NAOC). An Optomechanics Research Inc. (OMR) spectrograph and a PI 1340 $\times$ 400 CCD were used, which result in a two-pixel resolution of $\sim$ 2.0 \AA~. The spectral coverage of the observations is from 6300 to 7600 \AA~. To allow greater throughput for the weak emissions of faint nebulosity observed on this run, a slit width of 3.$\arcsec$0 was set through bright core of the source along the north-south (NS) direction. The diaphragm sizes of the slit can cover most parts of this object. The exposure times ranged from 1200 to 6900 s, resulting in signal-to-noise ratios (S/N) of $>$ 80. The seeing conditions during the observing run varied between 2.$\arcsec$2 and 2.$\arcsec$8. Exposures of He-Ar arcs were obtained right before and after each stellar spectrum and used for the wavelength calibration.

Data were reduced following standard procedures in the NOAO IRAF V2.16 software package. The CCD reductions included bias and flat-field correction, successful background subtraction and cosmic-ray removal. Flux calibration was derived from observations of at least three of the KPNO standard stars per night. The atmospheric extinction was corrected by the mean extinction coefficients measured for Xing-Long station, where the 2.16 m Telescope is located. A final spectrum for the nebula was produced using the combined OMR observations to improve the S/N and a summary of the spectroscopic observations is given in Table \ref{tab2}.

\begin{figure}
\begin{center}
\includegraphics[width=0.5\textwidth]{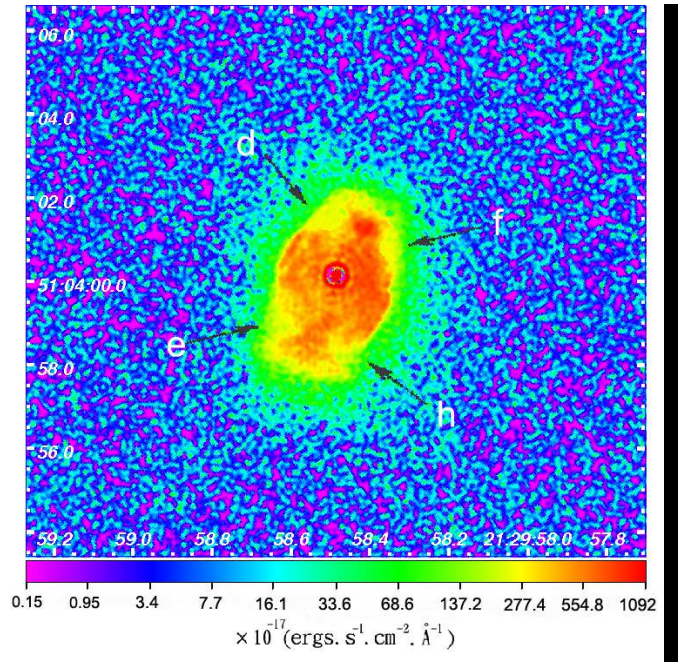}
\end{center}
\caption{{\it HST} WFPC2 H$\alpha$ image of IRAS 21282+5050. The main nebula seen in this H$\alpha$ image corresponds to the central cylindrical structure (marked as ``$g$'') in Figure~\ref{acs}. 
Several low-density regions are marked as $d, e, f$, and $h$. 
North is up and east is to the left. 
The intensity display is on a logarithmic scale and the grey scale bar is given at the bottom in units of ergs cm$^{-2}$ s$^{-1}$ \AA~$^{-1}$}.\label{ha}
\end{figure}

%\subsection{ISO spectra}

%Infrared spectra of IRAS 21282 were obtained by the {\it Infrared Space Observatory} (ISO) %ranged from 1996 January to 1996 November.
%using the Short Wavelength Spectrometer (SWS, PI:M. Jourdain de Muizon) module (2.4 - 45.2 $\mu$m, $\lambda$/$\triangle\lambda$ = 1600) and Long Wavelength Spectrometer (LWS, PI:M. Barlow) module (43 - 196.7 $\mu$m, $\lambda$/$\triangle\lambda$ = 200). The aperture sizes of SWS and LWS modules are 33$\arcsec\times$ 20$\arcsec$ and 84$\arcsec\times$ 84$\arcsec$, respectively, and therefore should cover entire nebula.         

%Data were reduced through the ISO Spectral Analysis Package (ISAP) to remove bad data points, flat-fielding the scans.  To improve the S/N of ISO observations, the final SWS and LWS spectra were performed using the combined data. Although the SWS and LWS observations of this object have been presented by \citet{Fong01} and \citet{Liu01}, the characteristics of broad emission features and dust components are still unclear and are needed to investigate. Here we re-process the spectra and try to draw out more properties from these components. The journal of ISO spectroscopic observations is given in Table \ref{tab3}.

\begin{figure*}
\begin{center}
\includegraphics[width=1\textwidth]{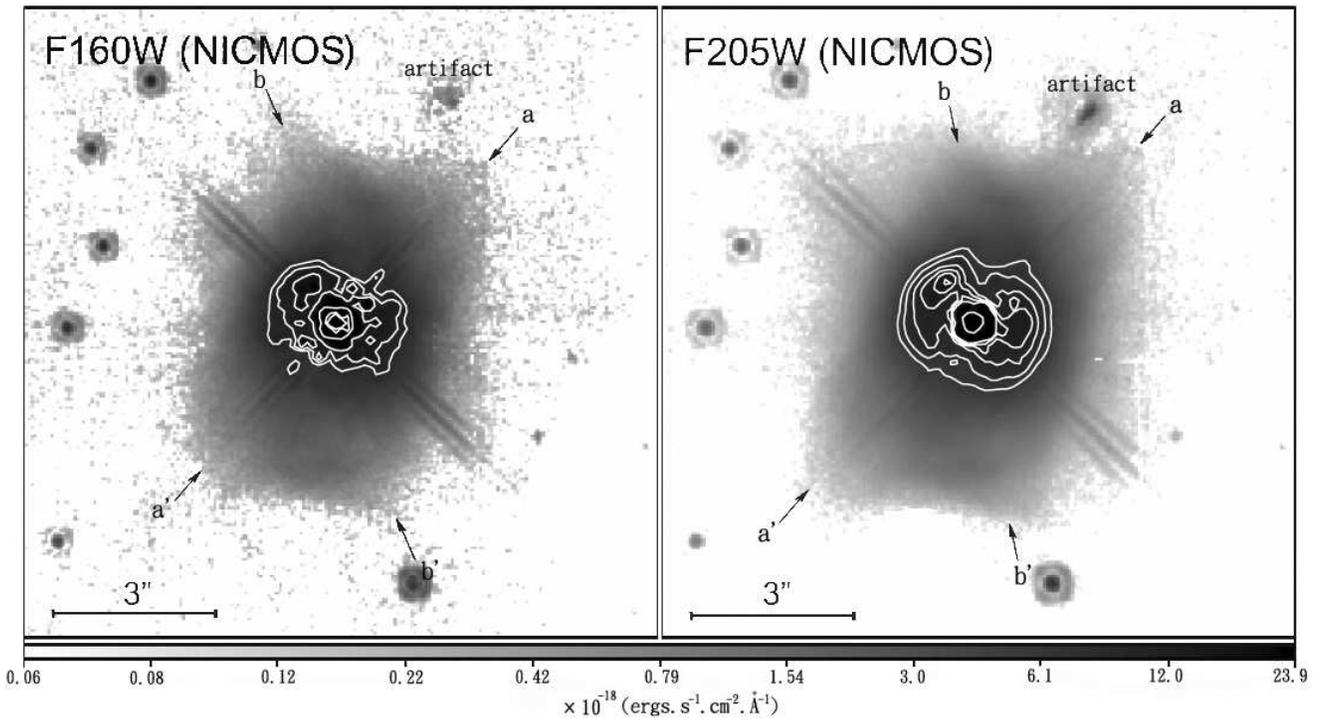}
\end{center}
\caption{{\it HST} NICMOS F160W (left) and F205W (right) images of IRAS 21282+5050. 
In the center is a bright torus (shown in logarithmic white line contours. The outer four contours are drawn at the 155, 200, 260, and 340 $\sigma$ level above background sky brightness. 
The central compact contours indicate the position of the central star. 
Two pairs of bipolar lobes ($a-a^\prime$ and $b-b^\prime$) can be seen in both images, but the lobe $c-c^\prime$ in Fig.~\ref{acs} is not found in the NICMOS images. 
North is up and east is to the left. 
The intensity display is on a logarithmic scale and the grey scale bar is given at the bottom in units of ergs cm$^{-2}$ s$^{-1}$ \AA~$^{-1}$.
\label{nicmos}}
\end{figure*}

\section{Results}

\subsection{Multiple lobes}\label{lobes}

Our {\it HST} ACS broad V-band (F606W) image of IRAS 21282 (Figure~\ref{acs}) clearly shows that this nebula is extended along the N-S direction with a size of $\sim$ 8.$\arcsec$51 $\times$ 6.$\arcsec$15, which is slightly larger than the size of 6.$\arcsec$5 $\times$4.$\arcsec$8 measured from the first optical image observed at {\it Canada-France-Hawaii Telescope} \citep{Kwok93}. 
Three pairs of bipolar lobes (labeled as $a-a^\prime$, $b-b^\prime$, and $c-c^\prime$) emanating from the center of this nebula can also be seen. 
At the center is an elliptical cylindrical  shell  (marked as $g$) along the approximate SSE-NNW direction (PA = 157 $^\circ\pm$ 5 $^\circ$). 
%This structure could represent an elliptical shell or alternatively could be the result of projection of the fourth pair of bipolar lobes aligned nearly along the plane of the sky. 
Another extended diffuse structure (labeled as ``torus'') can also be seen along the E-W direction around the waist of this nebula. 
%Also can be seen in this image are some extended outer nebulosities in the NNE direction.

The axes of these three pairs of lobes $a-a^\prime$, $b-b^\prime$, and $c-c^\prime$ intersect approximately at the position of the central star seen in Figs.~\ref{ha} and \ref{nicmos}. 
The position angles (PAs) of these three bipolar lobes are measured to be PA = 139$^\circ$ $\pm$ 3$^\circ$, 13$^\circ$ $\pm$ 2$^\circ$, and 35$^\circ$ $\pm$ 3$^\circ$ for lobes $a-a^\prime$, $b-b^\prime$, and $c-c^\prime$, respectively, which are not perpendicular to each other.  
Sizes of the lobes are measured by fitting the ellipses to the image. These three bipolar lobes have similar projected sizes on the sky. 
The measured angular sizes of three pairs of lobes ($a-a^\prime$, $b-b^\prime$, and $c-c^\prime$) are 9.$\arcsec$96 $\times$ 2.$\arcsec$88, 9.$\arcsec$02 $\times$ 1.$\arcsec$64, and 9.$\arcsec$66 $\times$ 3.$\arcsec$04.

Precise distances of PNs are extremely difficult to determine \citep[see e.g.,][]{sta17}. Based on a comparison between the systematical velocity and the Galactic rotation, \citet[]{lik88} suggested that IRAS 21282 is close to the Sun with a distance of $<2$\,kpc. However, \citet[]{shi89} found that in order to reach good agreement between the kinematical age of gas and the evolutionary age of the central star, one has to assume a distance of 2\,kpc. A larger distance value of 3.38\,kpc was derived from the four IRAS band fluxes, assuming a total infrared luminosity of 8,500 L$_\sun$ for compact PNs \citep[]{cas01}. The recent data release (DR) of Gaia trigonometric parallexes of a large sample of PNs has allowed the derivation of  model-independent distances for more PNs. \citet[]{sta17} and \citet[]{kim18} presented the distances of PNs in Gaia DR1 and DR2, respectively. The parallex of IRAS 21282+5050 was published in Gaia DR2. However,  \citet{Bailer18} argued that reliable distance cannot be obtained by inverting the parallax. Instead, they obtained distance scales by using a weak distance prior according to Galactic model, suggesting a distance of $3.57_{-0.22}^{+0.26}$\,kpc for IRAS 21282+5050. This more reliable distance allows us to evaluate distance-dependent parameters of this PN with higher accuracy.

Adopting the distance of 3.57 kpc for IRAS 21282 \citep{Bailer18}, the physical size of the total extent of lobe $a-a^\prime$ is $\sim$0.17 $\sec$$\theta$ pc , where $\theta$ is the inclination angle. 
Assuming the expansion velocity of 50 km s$^{-1}$ \citep{Cohen87}, we derive a kinematic age of $\sim$ 3400 $\sec$$\theta$ yr. This is consistent with the earlier suggestion that IRAS 21282 is a very young PN \citep{Kwok93}.

%If these lobes all lie close to the plane of the sky, the estimated kinematic age of this object is small. From above estimation, the peculiar nebula IRAS 21282 is young, which suggests that this nebula left the AGB stage only a short time ($<$ 10$^{3}$ years) ago \citep{Kwok93}. 
%The expansion velocities of a multipolar PN in the 
%lobes usually have larger values than that in other regions, thus the %derived kinematic age here is likely to be the 
%upper limit. 
%Because of inclination effects, the derived kinematic age can be considered as an approximate lower limit.
%Considering the uncertain determination of projected size of the PN caused by the inclination effect in a 2D projection model, the derived kinematic age presented here is just a roughly estimate.

The WFPC H$\alpha$ image (Figure~\ref{ha}) shows an elliptical cylinder shell surrounding a bright central star. This shell corresponds to the cylinder-shaped structure labeled $g$ in Figure~\ref{acs}. This elliptical cylinder structure has an angular size of 4.$\arcsec$92 $\times$ 3.$\arcsec$01.
The major axis of the structure is oriented at PA = 159$^\circ\pm$ 3$^\circ$, which is different from the axis of the lobe $a-a^\prime$. 
The nebula shows an edge with high density contrast as well as several clumpy regions (Figure~\ref{ha}). %The origin of these clumpy structures and the physical mechanisms leading to their formation are not clear. 
It is possible that these regions are the results of projection of interlaced multiple lobes and/or ionized non-uniform dense nebula emitted by the central star. 
Several low-density regions ($d, e, f, h$) associated with the lobes $a-a^\prime$ and $c-c^\prime$ are also found. The multiple lobes seen in IRAS 21282 may represent the directions where the UV photons are escaping.

\subsection{Dust torus}\label{torus}

The two broad-band near-infrared F160W and F205W images (Fig.~\ref{nicmos}) have similar appearances. The counterparts of bipolar pairs $a-a^\prime$ and $b-b^\prime$ can be seen in Figure~\ref{nicmos}, but the $c-c^\prime$ bipolar lobe is not. 
%A spatially resolved structure with a size of 6.$\arcsec$11 $\times$ 4.$\arcsec$13 can be found in the central region of IRAS 21282. 
The central region of the  nebula shows an elliptically shaped structure of size $\sim$6.$\arcsec$11 $\times$ 4.$\arcsec$13 with two prominent peaks along the ENE-WSW direction.  
The symmetry axis of the infrared nebula lies at the PA of 155$^\circ$ $\pm$ 2$^\circ$, which is almost the same orientation as the elliptical shell structure observed in the H$\alpha$ image (see Section \ref{lobes}). 
The nebula is therefore likely to be an infrared counterpart of the H$\alpha$ cylinder-shaped shell. 
These two bright peaks (separated by 1.$\arcsec$65) lie along an axis with PA = 63$^\circ$ $\pm$ 3$^\circ$, 
which is perpendicular to the major axis (polar direction) of the nebula. 

If the double-peaked structure seen in the F160W and F205W images represent an oblique torus, then the emission from these two peaks could be the result of projection of a toroid viewed edge-on, implying that the elliptical nebula is oriented close to the plane of the sky. Such toroidal structures can be found in the PNs with binary nuclei  \citep[e.g., NGC 2346,][]{Su04} and  \citep[M2-9,][]{Castro10}, and has been interpreted as the result of sweeping up previously-ejected AGB circumstellar materials by a later developed fast, collimated wind \citep{Mastrodemos98}.

\begin{figure*}
	\begin{center}
		\includegraphics[width=0.9\textwidth]{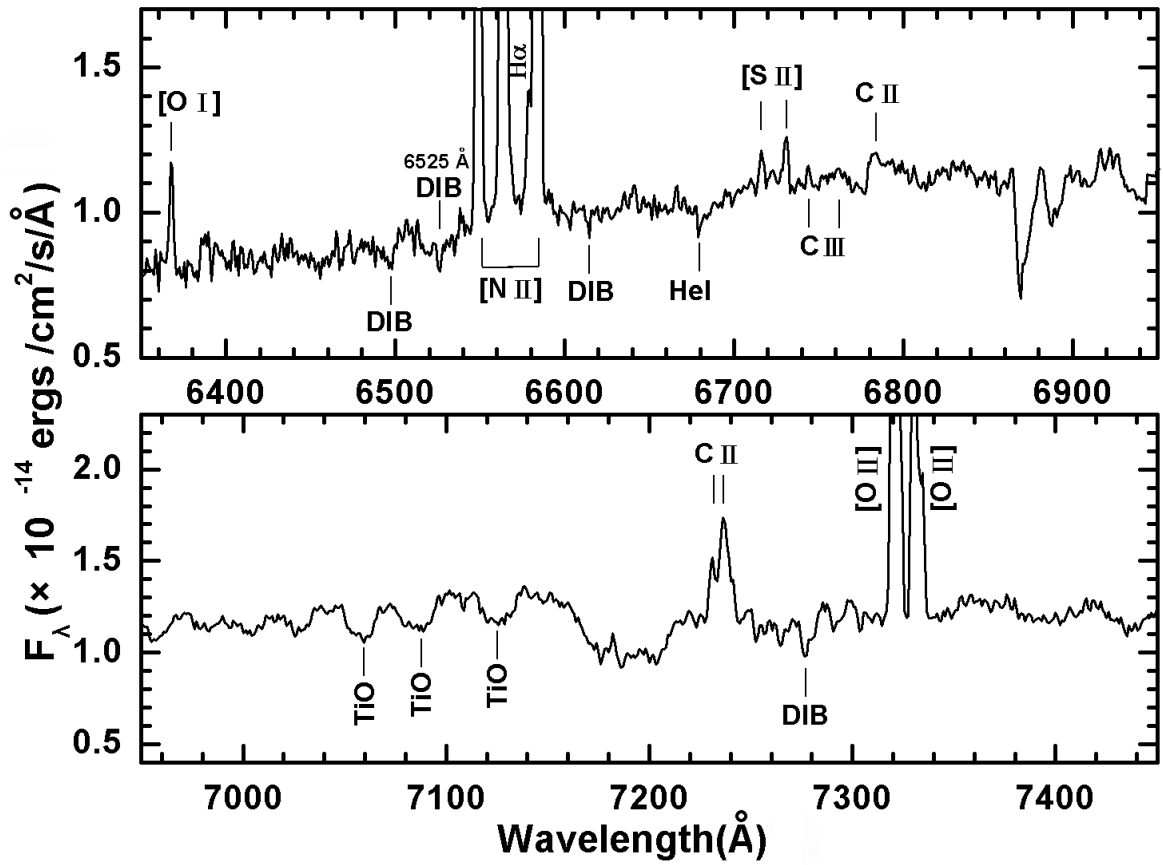}
	\end{center}
	\caption{NAOC OMR spectrum of IRAS 21282+5050 (not corrected for extinction) in wavelength range of 6350 \AA~ to 7450 \AA~. The 
emission lines, DIBs, and absorption lines are marked.}
	\label{omr}
\end{figure*}

This torus is not obvious in the H$\alpha$ image, probably because the dust torus is optically thick to the UV photons and the this structure is not ionized. 
The near-infrared morphology of IRAS 21282 (Figure~\ref{nicmos}) seems to be consistent with that at H-band \citep{Cheng05}, 2.2 $\mu$m \citep{Meixner93}, and H$_{2}$ images \citep{Davis05}, all of which show a central peak, probably the result of diffuse scattering light from the central star or as the result of the low resolution of these images. 
Further comparisons between our {\it HST NICMOS} observations with the deconvolved infrared images at 3.3, 8.5, 10.0, 11.3, and 12.5 $\mu$m of \citet{Meixner93} and Keck ones at K-continuum, L$^\prime$, Ms, and PAH bands of \citet{Cheng05} show that the elongation direction of this nebula and the locations of two bright peaks are very similar, which indicates that the infrared emission in these images are mainly dominated by the same dust continuum or some contributions from UIE bands \citep{Meixner93}. The dust component heated by the starlight escapes more directly through 
the polar direction. The presence of dust torus in IRAS 21282 also suggests that the mass of the waist part is much larger than that of its poles.
 
\begin{table*}
	\caption{Atomic lines in IRAS 21282+5050}
	\centering
	\begin{tabular}{lccccr}
		\hline\hline
		& \multicolumn{2}{c}{Identification} & & & \\
		\cline{2-3}
		$\lambda_{obs}$ & $\lambda_{lab}$ & Ion & Observed Flux$^a$ & Dereddened Flux$^a$ & RV$^b$ \\
		(\AA~)  &  (\AA~) &  &   &  & (km~s$^{-1}$) \\
		\hline
	    \multicolumn{6}{c}{Emissions} \\
		\hline	
	6364.12 & 6363.78 & [O{\sc i}] & 1.25(23.2) & 1.50(23.2) & 16 $\pm$ 6 \\
	6549.15 & 6548.10 & [N{\sc ii}] & 38.42(1.4) & 38.94(1.4) & 48 $\pm$ 3 \\
        6563.75 & 6562.77 & H$\alpha$ & 100.00(0.9) & 100.00(0.9) & 45 $\pm$ 2 \\ 
 	6577.55$^c$ & 6578.03 & C{\sc ii} & 0.40(50.5) &  0.39(50.9) & -22 $\pm$ 11 \\
	6584.20 & 6583.50 & [N{\sc ii}] & 125.00(2.5) &  122.68(2.5) & 32 $\pm$ 3 \\
	6716.65 & 6716.40 & [S{\sc ii}] & 0.80(19.6) & 0.70(19.6) & 11 $\pm$ 12 \\
	6722.21 & 6721.35 & [O{\sc ii}] & 0.30(58.4) &  0.26(58.4) & 38 $\pm$ 19 \\
	6730.91 & 6730.82 & [S{\sc ii}] & 1.11(12.7) & 0.96(12.7) & 4 $\pm$ 9 \\
	6743.93 & 6744.40 & C{\sc iii} & 0.05(36.1) & 0.04(36.1) & -21 $\pm$ 12 \\
	6761.42 & 6762.20 & C{\sc iii} &  0.04(48.4) &  0.04(48.5) & -35 $\pm$ 18 \\
	6779.66$^d$ & 6779.90 & C{\sc ii} & 0.53(6.8) &  0.44(6.8) & -11 $\pm$ 8 \\
	6782.74$^d$ & 6783.90 & C{\sc ii} & 0.22(19.1) & 0.18(19.1) & -51 $\pm$ 19 \\
	6798.28 & 6798.10 & C{\sc ii} & 0.50(43.2) & 0.41(43.2) & 8 $\pm$ 15 \\
	7230.48 & 7231.30 & C{\sc ii} & 0.90(11.8) & 0.52(11.8) & -34 $\pm$ 6 \\
	7236.45 & 7236.40 & C{\sc ii} & 1.41(9.8) & 0.81(9.9) & 2 $\pm$ 5 \\
	7321.15 & 7319.99 & [O{\sc ii}] & 24.47(10.9) & 13.19(10.9) & 48 $\pm$ 5 \\
	7331.98 & 7330.73 & [O{\sc ii}] & 19.33(17.1) & 10.34(17.1) & 51 $\pm$ 5 \\
		\hline
		\multicolumn{6}{c}{Photospheric absorption from hot star} \\
		\hline	
		6678.84 & 6678.15 & He{\sc i} & ... & ... & 31 $\pm$ 22 \\
		\hline
	\end{tabular}
%	\tablefoot{
		\tablenotetext{a}{The line fluxes were normalized to F(H$\alpha$)=100 and percentage uncertain errors of the flux measurements are given within brackets.}
		\tablenotetext{b}{Error measured using the method described in \citet{Manick15}.}
		\tablenotetext{c}{Originated from the companion star.}
		\tablenotetext{d}{From Gaussian fitting routine of closed lines.}
%	}
	\label{tab4}
\end{table*} 

\subsection{Optical spectroscopy}\label{spectra}

Although optical spectra with different resolutions of this nebula have been presented by \citet{Cohen87} and \citet{Leuenhagen98}, their studies focused on the spectral type of central star, velocity distributions of lines, and element abundances of nitrogen, neon, and silicon of the nebula. 
We have performed new spectroscopic observations and the results are shown in Figure~\ref{omr}. The spectrum of IRAS 21282 is dominated by prominent emission lines of [N{\sc ii}] $\lambda\lambda$6548, 6584,  H$\alpha$, [S{\sc ii}] $\lambda\lambda$6717, 6731, and [O{\sc ii}] $\lambda\lambda$7319, 7330. Some weak emission features such as C{\sc ii} $\lambda\lambda$6578, 6779, 6784, 6798, 7231, 7236 and C{\sc iii} $\lambda\lambda$6744, 6762 can also be seen. The C{\sc ii}/C{\sc iii} emission ratio suggests that the nebula is carbon-rich with a cool [WC 11] nucleus \citep{Cohen87, van81}. From the flatness of the quotient spectrum through IRAS 21282 divided by a reddened O7- type star (see Figure 3, Cohen \& Jones 1987), an extinction coefficient value of A$_{v}$ = 5.8 mag was derived. This implies an E(B-V) value of 1.88, which can be attributed mostly to the circumstellar extinction.

The emission line fluxes and their radial velocities (RVs) measured from the spectrum are listed in Table~\ref{tab4}. The measured central wavelengths of the emissions and line identifications are listed in Columns 1 - 3. Column 4 and 5 give the observed and dereddened fluxes (normalized to H$\alpha$ = 100) measured using the Gaussian fitting routine for this nebula. The RVs of the line features measured from our observation are given in Column 6. Although the H$\alpha$ line measurements are affected by central-star absorption, the flux uncertainties due to the effect are relatively minor. The raw integrated H$\alpha$ flux measured from main nebula for this object is 3.24 $\times$ 10$^{-13}$ erg cm$^{-2}$ s$^{-1}$. The characteristic uncertainty of flux measurements is about 24$\%$. The line ratio of [N{\sc ii}] $\lambda$6584/$\lambda$6548 of IRAS 21282 is 3.2, in good agreement with the theoretical predictions. 
Using the [S{\sc ii}] $\lambda$6731/$\lambda$6717 line ratio and assuming an electron temperature $T_{e}$ = 10,000 K, we derive the electron density $n_{e}$ = 1900 cm$^{-3}$ for the nebula, which is consistent with the previously reported range of 
2,000 -- 10$^{4}$ cm$^{-3}$ \citep{Likkel94} and 200 -- 10$^{5}$ cm$^{-3}$ \citep{Cohen87}.

\subsubsection{Circumstellar diffuse interstellar bands}
One of the unique aspect of IRAS 21282 is the presence of circumstellar diffuse interstellar bands (DIB). 
Previously, DIBs at $\lambda$$\lambda$5780, 5797, 6177, 6203, 6270, and 6613 \AA~ have been detected by \citet{Cohen87} and \citet{Le92}. In Figure~\ref{omr}, we show the detections of DIBs at $\lambda$$\lambda$6498, 6613, and 7276 \AA. 
In addition, a new DIB band at 6525 \AA~ \citep[first discovered in PN Tc 1,][]{Garcia13} is also detected in our spectrum. This feature may be related to larger fullerenes (e.g., C$_{80}$, C$_{240}$, C$_{320}$, and C$_{540}$) and buckyonions in the circumstellar envelope. 

\subsubsection{A cool companion}

Interestingly, molecular TiO absorptions at $\lambda\lambda$7054, 7088, 7126 \AA~ characteristic of late-type stars can be seen in the spectrum. These TiO absorption bands suggest the presence of a cool companion to the exciting source of this nebula, raising the possibility that there may be a binary nucleus in IRAS 21282 \citep{Cohen87, Meixner93}. 

%As shown in Table~\ref{tab4}, photospheric absorption of O-type star (He{\sc i} $\lambda\lambda$6678 \AA) has a velocity of 31 $\pm$ 22 km~s$^{-1}$ and C{\sc ii} emission at $\lambda$$\lambda$6578 \AA~ originated from secondary \citep{Hillwig15} with a velocity of -22 $\pm$ 11 km~s$^{-1}$. Although we can not gauge systemic velocity of this object directly, the velocity difference for exciting star between the hot star and cool companion can serve as a probe of symbiotic system and be indicative of binary nucleus of the nebula. New high-resolution spectroscopic observations are needed to further determine the nature of the cool companion.

\begin{table*}
	\caption{Photometric measurements of IRAS 21282+5050}
	\centering
	\begin{tabular}{llc}
		\hline\hline
		Filters & Flux/Flux density & Reference \\
		\hline
		\multicolumn{3}{c}{Central star and nebula} \\
		\hline
		U (mag) & 16.33 & \citet{Cohen87} \\
		B (mag) & 15.98 & \citet{Cohen87}  \\
		V (mag) & 14.40 & \citet{Cohen87} \\
		R (mag) & 13.39 & \citet{Cohen87} \\
		I (mag) & 12.57$\pm$0.02 & \citet{Kwok93} \\
		%-log F (H$\alpha$) (ergs.cm$^{-2}$.s$^{-1}$)$^c$ & 12.49 \\
		\hline
		\multicolumn{3}{c}{Dust} \\
		\hline
		L$^\prime$ (mag) & 6.59$\pm$0.02 & \citet{Kwok93} \\
		M (mag) & 5.80$\pm$0.04 & \citet{Kwok93} \\
		2MASS J (mag) & 11.504$\pm$0.019 & \citet{Cutri03} \\
		2MASS H (mag) & 10.709$\pm$0.017 & \citet{Cutri03} \\
		2MASS Ks (mag) & 9.591$\pm$0.020 & \citet{Cutri03} \\
		WISE F 3.4 $\mu$m (Jy) & 0.61$\pm$0.02 & \citet{Cutri12} \\ 
		WISE F 4.6 $\mu$m (Jy) & 0.94$\pm$0.02 & \citet{Cutri12} \\
		WISE F 12 $\mu$m (Jy) & 77.95$\pm$1.46 & \citet{Cutri12} \\    
		WISE F 22 $\mu$m (Jy) & 62.10$\pm$0.10 & \citet{Cutri12} \\
		MSX F 8.3 $\mu$m (Jy) & 24.58$\pm$1.01 & \citet{Egan03} \\
		MSX F 12.1 $\mu$m (Jy) & 62.05$\pm$3.10 & \citet{Egan03} \\
		MSX F 14.7 $\mu$m (Jy) & 67.08$\pm$4.11 & \citet{Egan03} \\
		MSX F 21.3 $\mu$m (Jy) & 61.79$\pm$3.71 & \citet{Egan03} \\
		IRAS F 12 $\mu$m (Jy) & 50.54$\pm$1.77 & \citet{Moshir92} \\
		IRAS F 25 $\mu$m (Jy) & 72.82$\pm$2.99 & \citet{Moshir92} \\
		IRAS F 60 $\mu$m (Jy) & 31.88$\pm$1.28 & \citet{Moshir92} \\
		IRAS F 100 $\mu$m$^a$ (Jy) & 31.81$\pm$2.99 & \citet{Moshir92} \\
		AKARI F 9 $\mu$m (Jy) & 26.51$\pm$0.09 & \citet{Ishihara10} \\
		AKARI F 18 $\mu$m (Jy) & 64.05$\pm$0.47 & \citet{Ishihara10} \\
		AKARI F 65 $\mu$m (Jy) & 25.73$\pm$1.10 & \citet{Ishihara10} \\
		AKARI F 90 $\mu$m (Jy) & 19.62$\pm$0.77 & \citet{Ishihara10} \\
		AKARI F 140 $\mu$m (Jy) & 5.03$\pm$0.51 & \citet{Ishihara10} \\
		AKARI F 160 $\mu$m$^b$ (Jy) & 3.71$\pm$2.71 & \citet{Ishihara10} \\
		\hline
		\multicolumn{3}{c}{Free-free emission} \\
		\hline
		IRAM F 1300 $\mu$m (mJy) & 5.40$\pm$1.20 & \citet{Castro10} \\
		IRAM F 2600 $\mu$m (mJy) & 2.90$\pm$0.60 & \citet{Castro10} \\
		15 GHz (mJy) & 6.80$\pm$1.02 & \citet{Likkel94}\\ 
		8.4 GHz (mJy) & 4.28$\pm$0.04 & \citet{Knapp95} \\
		5 GHz (mJy) & 6.90$\pm$0.69 & \citet{Likkel94} \\
		5 GHz (mJy) & 8.00$\pm$0.80 & \citet{Meixner93} \\
		\hline
	\end{tabular}
%	\tablefoot{
		\tablenotetext{a}{Note that the flux measured from 100 $\mu$m is an upper limit.}
		\tablenotetext{b}{From AKARI database. Note that the 160 $\mu$m flux is uncertain.}
%	}
	\label{tab5}
\end{table*}

\begin{table*}
	\caption{{\it ISO} observations of IRAS 21282}
	\centering
	\begin{tabular}{clcccc}
		\hline\hline
		PN & Name & Instrument & Observation Date & Wavelength Range & Exposures \\
		\hline
		&      & \multicolumn{2}{c}{SWS Spectra} & & \\
		\hline
		G 093.9-00.1 & IRAS 21282+5050 & TDT 05602477 & 1996 Jan 12 & 2.4 - 45.2 $\mu$m & 1860 s  \\
		& & TDT 15901777 & 1996 Apr 24 & 2.4 - 45.2 $\mu$m & 3462 s\\
		& & TDT 36801940 & 1996 Nov 18 & 2.4 - 45.2 $\mu$m & 1913 s\\
		\hline
		& & \multicolumn{2}{c}{LWS Spectra} & & \\
		\hline
		G 093.9-00.1 & IRAS 21282+5050 & TDT 36801941 & 1996 Nov 18 & 43 - 186.5 $\mu$m & 1004 s \\
		& & TDT 34602320 & 1996 Oct 28 & 43 - 195.8 $\mu$m & 1268 s \\
		\hline
	\end{tabular}
	\label{tab3}
\end{table*}

\section{Spectral energy distribution}

\begin{figure}
        \begin{center}
                \includegraphics[width=0.5\textwidth]{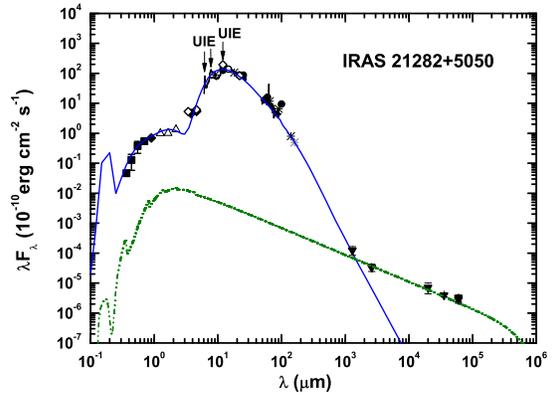}
        \end{center}
        \caption{SED of IRAS 21282+5050 in the wavelength range from 0.1 $\mu$m to 1 m. The U, B, V and R photometry are shown as filled squares, the I, L$^\prime$ and M near-IR photometry as filled diamonds, the 2MASS photometry as open triangles, {\it WISE} photometry as open diamonds, {\it MSX} as open circles, {\it IRAS} as filled circles, and {\it AKARI} photometry as asterisks. 
The light asterisk represents the uncertain {\it AKARI} detection. 
The {\it IRAS} 100 $\mu$m flux is an upper limit. 
The {\it ISO} SWS and LWS spectra are also plotted. 
The green curve represents the nebular gas {\it b-f} and {\it f-f} emissions. 
The blue curve is a dust continuum radiation transfer model (DUSTY \citep{Ivezic99}) fit of the emergent flux. 
The excess in the ultraviolet part of the model curve is the reddened photosphere of the hot central star.}
        \label{sed}
\end{figure}

Young PNs often show strong infrared excess due to thermal emission of dust components and a significant fraction of their total energy output is emitted in the infrared \citep{Zhang91}. In order to determine the relative contributions of the photospheric, nebular gas, and dust components to the total observed fluxes for young PNs, we have constructed spectral energy distributions (SEDs) for IRAS 21282 (Figure~\ref{sed}). The optical U, B, V, and R photometry of central star of IRAS 21282 are obtained from \citet{Cohen87}. Near-infrared photometry at I, $L^\prime$, M, J, H, and K bands from ground-based and  {\it Two Micron All Sky Survey} (2MASS) observations are taken from \citet{Kwok93} and 2MASS database, respectively. In the mid-infrared, data from {\it Midcourse Space Experiment} (MSX), {\it IRAS}, and {\it AKARI} Point Source Catalogues are used. We have also added photometric measurements derived from the infrared images form {\it Wide-field Infrared Survey Explorer} (WISE; Wright et al. 2010) survey. Also plotted on Figure~\ref{sed} are the {\it Infrared Space Observatory} ({\it ISO}) SWS and LWS observations of the object. 
Summaries of the photometric and spectroscopic  data used are given in Tables \ref{tab5} and \ref{tab3}, respectively\footnote{We note that some filters are broad and color corrections may be needed for some flux densities given in Figure~\ref{sed} and Table \ref{tab5}. The color correction factors for the {\it WISE} bands can be found in \citet{Wright10}}.

\subsection{Total flux emitted by IRAS 21282}

We can see from the SED that most of the flux from IRAS 21282 is emitted in the mid-infrared. If we take the peak flux ($\lambda$$F_{\lambda}$) of the SED as $\sim$ 10$^{-8}$ erg cm$^{-2}$ s$^{-1}$ and assume that this is the peak of a blackbody, then the total emitted flux is approximately ($\lambda$$F_{\lambda}$)$_{peak}$/0.736 = 1.4 $\times$ 10$^{-8}$ erg cm$^{-2}$ s$^{-1}$ (Kwok 2007, p. 521). 
Assuming a distance of 3.57 kpc, the total luminosity emitted by the object is 5600 L$_{\sun}$. Since the observed SED is broader than a single-temperature blackbody, this value should be a minimum. If we plot the SED in units of $F_{\lambda}$ and sum up the area under the curve of the SED, the total flux is 1.6 $\times$ 10$^{-8}$ erg cm$^{-2}$ s$^{-1}$, giving a luminosity of $\sim$ 6400 L$_{\sun}$.

\subsection{Evidence for a cool companion}

The observed optical fluxes represent the sum of the reddened photospheric emission from the central star and nebular emissions. The gaseous nebular continuum is the sum of the bound-free ($b-f$), free-free ($f-f$), and two-photon emissions. We use the emission coefficients given in \citet{Kwok07} at a typical electron temperature $T_{e}$ = 10$^{4}$ K. At wavelengths longward of $\lambda$ 2 mm, $f-f$ emission dominates over dust emission. The emission measure of the nebular gas component can therefore be constrained by the observed $f-f$ fluxes in the radio.

From the SED, we note that the observed {\it WISE} 3.4 $\mu$m and $L^\prime$-band fluxes are higher than the expected near-infrared continuum. These excesses could be due to contribution from the 3.3 $\mu$m UIE band \citep{Meixner93, Hora99}. UIE bands at 6.2, 7.7-7.9, 8.6, 11.3, 12.7, and 14.2 $\mu$m and a weak  band at 6.0 $\mu$m be seen in the {\it ISO} spectrum \citep{Hsia16}.

%The mid- to far-infrared component of the SED is too broad to be fitted by a single dust temperature. 
%We have tried to fit the observed SED by a radiation transfer model using the software code DUSTY \citep{Ivezic99}. For the fitting, we assumed the standard MRN distribution of grain sizes \citep{Mathis77} and a density distribution of R$^{-2}$, where R is the distance to the central star. In order to arrive at the best fitting, the optical thickness of the shell at 0.55 $\mu$m was adopted to be $\tau$ = 5.5. The dust grain is assumed to be amorphous carbon, which will not fit the AIB features but can give an approximation to the dust continuum. 

The near-infrared continuum of the object cannot be fitted by the nebular gas component alone and must contain contributions from photospheric and/or hot dust components. 
In order to fit the optical and near infrared parts of the SED, we need two central stars of temperatures of 6,000 K and 25,000 K. Assuming a distance of 3.57 kpc, the respective luminosities of the two central stars are 1,900 and 5,400 L$_{\sun}$. The total luminosity of 
the object (7,300 L$_{\sun}$) is higher than the earlier estimate of 6,400  L$_{\sun}$ because part of the radiation from the central star in the UV is not represented in the total observed fluxes. The near infrared (from 0.8 to 5 $\mu$m) emission from IRAS 21282 are due to the cool companion as well as emission from the dust torus. Most of the infrared flux ($\lambda >5 \mu$m) must originate from a much larger cool 
dust envelope extending far beyond the optical nebula.

\begin{figure}
        \begin{center}
                \includegraphics[width=0.47\textwidth]{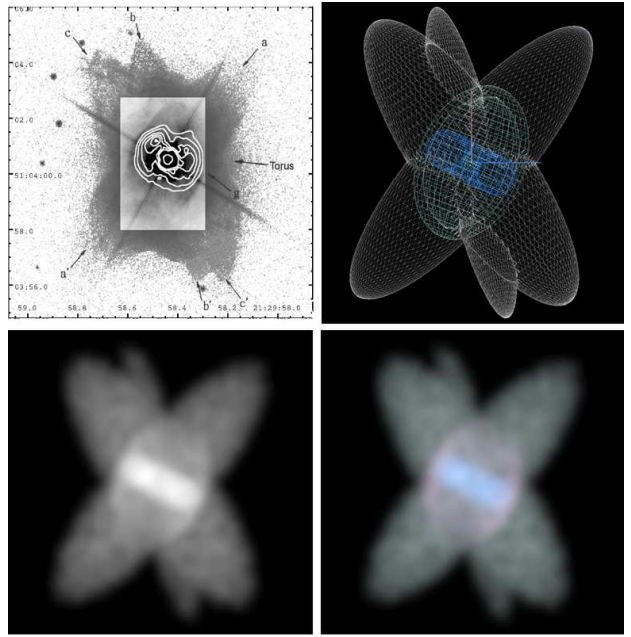}
        \end{center}
        \caption{Comparison of optical-infrared observations and the corresponding 3-D model images of IRAS 21282+5050. 
Upper left: {\it HST/ACS} F606W image (grey scale) is overlad with the NICMOS F205W image (shown as white-line logarithmic contours). The outer four contours drawn are the same as the plotting in Figure~\ref{nicmos}. 
Upper right: SHAPE 3-D mesh model. The bipolar lobes $a-a^\prime$, $b-b^\prime$, and $c-c^\prime$ are displayed in white and the fourth pair of lobes observed in the {\it HST} H$\alpha$ image (marked as $g$ in Figure~\ref{acs}) is shown in light blue. The infrared torus observed in {\it HST NICMOS} image is shown in dark blue. Lower left: the rendered image in gray scale. Lower right: rendered image with the outer three lobes ($a-a^\prime$, $b-b^\prime$, and $c-c^\prime$), inner bipolar lobe, and the equatorial torus shown in white, red, and blue, respectively. A Gaussian 
blur has been applied to the renderings to simulate the telescope beam.}
        \label{contour}
\end{figure}

\section{3-D model simulations}

From the results presented in Section \ref{lobes}, and \ref{torus}, we suggest that IRAS 21282 is a young multipolar PN. 
In order to visualize  its intrinsic structures, we have constructed a 3-D modeling using the software package \textit{SHAPE}\footnote{http://www.astrosen.unam.mx/shape/} \citep{steffen11}, which is a morpho-kinematic modeling tool used to analyze the 3-D structures of gaseous nebulae. 
%In this work, we are interested in studying the 3D geometry of multipolar PNs via SHAPE and comparing the rendered model images with observed ones. 
%We do not attempt to determine physical parameters such as temperature structure, wind velocity, and density distribution, and thus no hydrodynamic or radiative transfer simulation is performed. 
A 3-D multi-component model is constructed and the integrated emission intensities ($\int n_e^2 d\ell$) along the line of sight are compared with the observed surface brightness distribution in the {\it HST} images. 
%are assumed for the surface brightness of simulated images and excitation process is not considered in our model. 

Our model of IRAS 21282 consists of five major components: four pairs of bipolar lobes and an equatorial dust torus. 
%{\bf To model three pairs of bipolar lobes ($a-a^\prime$, $b-b^\prime$, and $c-c^\prime$) seen in Fig. \ref{acs}, it has been assumed 
%that the differences in their apparent lengths are attributed to the projection effect. 
The observed three pairs of bipolar lobes ($a-a^\prime$, $b-b^\prime$, and $c-c^\prime$) are directly incorporated in the model. We also create another small pair of lobes with openings at the tips that is aligned along the plane of the sky. This pair of ionized bipolar lobes is identified as a distinct component since it can be seen in the H$\alpha$ image (Figure~\ref{ha}), whereas the outer three lobes are only visible in the F606W broad-band image (Figure~\ref{acs}). 
The inner bipolar component with a shell-like appearance is referred to as lobe $g$ in Figure~\ref{acs} (see Section~\ref{lobes}). Assuming that the lengths of these three pairs of lobes are equal and lobe $a-a^\prime$ is perpendicular to the line of sight (at inclination angle $\theta=0^\circ$ relative to the plane of the sky), we derive the inclination angles of lobes $b-b^\prime$ and $c-c^\prime$ to be $-25^\circ$ and $-14^\circ$, respectively.

In addition, we also include a toroidal structure as the counterpart of infrared torus seen in the {\it HST NICMOS} images. The center of this dust torus is set to the intersection of the four pairs of bipolar lobes. The orientation of dust torus is fitted to the locations of two bright peaks, which have a separation distance of about 1.$\arcsec$16. According to the spectroscopic measurements at H$_{2}$ 2.12 $\mu$m \citep{Davis05} and our previous results, the inclination angle of observed torus is assumed to be 0$^\circ$ (viewed edge-on). 
In our model, the bipolar lobe $g$ is viewed side-on and its symmetry axis is perpendicular to the dust torus. After an inspection of the F606W and NICMOS images, three outer bipolar lobes ($a-a^\prime$, $b-b^\prime$, and $c-c^\prime$) and the dust torus of the nebula show no shell-like structure, these components are given a uniform density distribution. 
For the fourth pair of lobe seen in the H$\alpha$ image (assumed to correspond to ``$g$'' in Figure~\ref{acs}), the central element has a bipolar shell structure and is assumed to be identical. The observed and model parameters of these structures are listed in Table \ref{tab6}.

\begin{table*}
	\caption{Comparison of observed and model parameters of the lobes and torus}
	\centering
	\begin{tabular}{lccccccc}
		\hline\hline
		& \multicolumn{3}{c}{Observed} & & \multicolumn{3}{c}{Model} \\
		\cline{2-4}\cline{6-8}
		Parameters & Position angle$^a$ & Size$^b$ & Inclination angle$^c$ && Position angle$^a$ & Size$^b$ & Inclination angle$^c$ \\
		&        ($^\circ$) & ($\arcsec$) & ($^\circ$) & & ($^\circ$) & ($\arcsec$) & ($^\circ$) \\
		\hline
		Lobe $a-a^\prime$ & 139$\pm$3 & 9.96 $\times$ 2.88 & ... & & 139 & 9.96 $\times$ 2.88 & 0 \\
		Lobe $b-b^\prime$ & 13$\pm$2 & 9.02 $\times$ 1.64 & ... & & 13 & 9.96 $\times$ 1.64 & -25 \\
		Lobe $c-c^\prime$ & 35$\pm$3 & 9.66 $\times$ 3.04 & ... & & 35 & 9.96 $\times$ 3.04 & -14 \\
		Lobe g & 157$\pm$5 & 4.92 $\times$ 3.01 & 0$^d$ & & 157 & 4.48 $\times$ 3.01 & 0 \\
		Dust torus & 63$\pm$3$^e$ & 1.65$^f$ & 90$^g$ & & 63 & 1.65$^f$ & 90 \\
		\hline
	\end{tabular}
%	\tablefoot{
		\tablenotetext{a}{Measured from the major axis orientation.}
		\tablenotetext{b}{Measured by fitting the ellipses to the image.}
		\tablenotetext{c}{Assuming the orientation angle of sky plane is 0$^\circ$.}
		\tablenotetext{d}{From Davis et al. (2005).}
		\tablenotetext{e}{Derived from peak position of torus.}
		\tablenotetext{f}{Peak separation of torus.}
		\tablenotetext{g}{From Davis et al. (2005).}
%	}
	\label{tab6}
\end{table*}

A comparison of the {\it HST} optical-infrared observations and corresponding model images are shown in Figure~\ref{contour}, in which the model intensity distributions are convolved with a Gaussian to simulate the telescope beam. As we can see from the rendered images in Figure~\ref{contour} (lower panels), our modeling gives a good approximation to observed images of IRAS 21282 and some basic observed features can be reasonably reproduced by SHAPE model.

\begin{figure*}
        \begin{center}
                \includegraphics[width=1.0\textwidth]{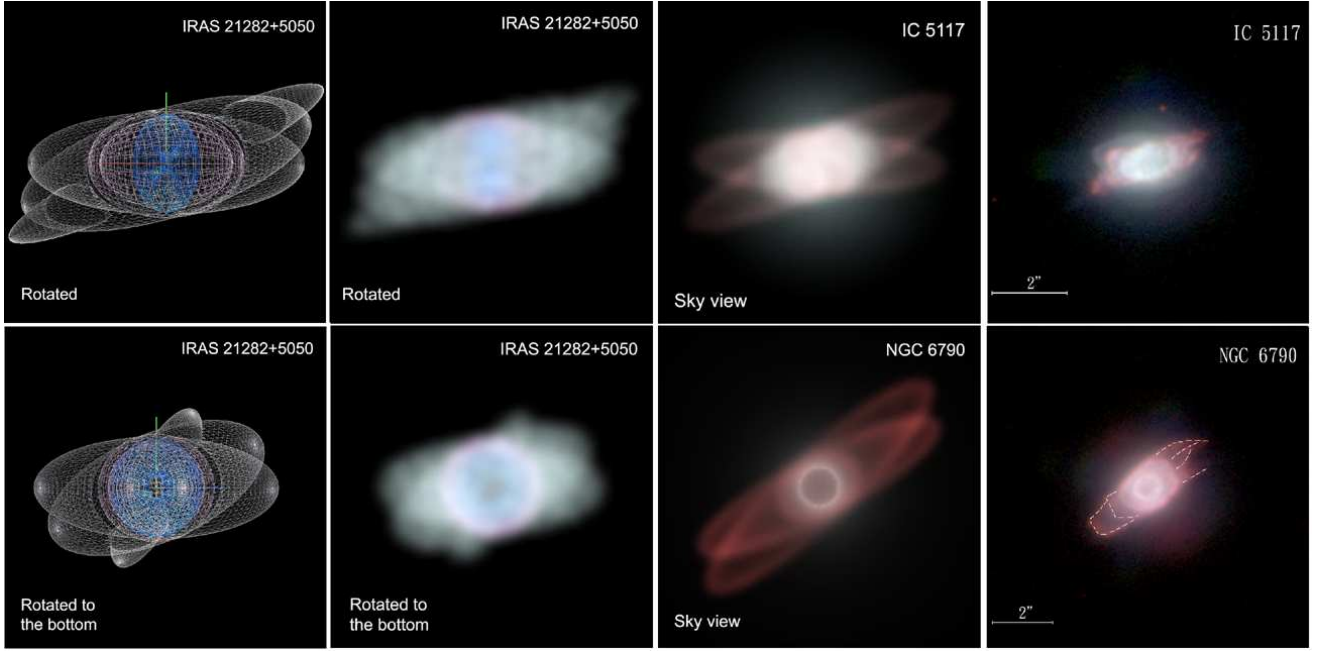}
        \end{center}
        \caption{Comparison of the model renderings of IRAS 21282 at different orientations and the model and observed images of multipolar PNs IC 5117 and NGC 6790. Upper panels, from left to right: (1) 3-D mesh image of IRAS 21282 rotated to the right and 90$^\circ$ counterclockwise; (2) the corresponding rendered image; (3) the rendered image of PN IC 5117; and (4) a composite-color {\it HST} image of IC 5117. Lower panels, from left to right: (1) 3-D mesh image of IRAS 21282 viewed from the bottom (near pole-on); (2) the corresponding rendered image; (3) the rendered image of PN NGC 6790; and (4) composite-color HST narrow-band image of NGC 6790. The {\it HST} and their
corresponding rendered images of IC 5117 and NGC 6790 are taken from \citet{Hsia14}.}
        \label{model}
\end{figure*}

Our model not only helps us to visualize  the  intrinsic 3-D structures of IRAS 21282 but can also predict  what this object may look like when viewed from different orientations. 
In Figure~\ref{model}, we show a comparison of the projected model images of IRAS 21282 viewed from different angles and compare these results to the images of two other young multipolar PNs (IC 5117 and NGC 6790). 
From Figure~\ref{model}, we note that the rendered images of IRAS 21282 differ from those of IC 5117 and NGC 6790. The bipolar lobes of IC 5117 and NGC 6790 show shell-like structures, they are modeled using identical bipolar shells, whereas the three pairs of bipolar lobes are constructed using a uniform density distribution for IRAS 21282. When the model is rotated to the right and next rotated 90$^\circ$ counterclockwise, the three pairs of lobes appear closer together and the nebula resembles a central elliptical shell. 
The appearance of rotated model is similar to the young PN IC 5117 \citep{Hsia14}. When the model is viewed from the bottom (pole-on), the object shows two pairs of distinct lobes and a bright ring, similar to the apparent structures of NGC 6790 \citep{Hsia14} and NGC 6644 \citep{Hsia10}. 
This suggests that one single multipolar structure may have different apparent morphologies when viewed from different orientations. The statement that ``each multipolar PN is unique'' is therefore unlikely to be true as different 2-D morphologies may originate from the similar intrinsic 3-D structures \citep{chong2012}. 

In the present exercise, we try to construct a 3-D model from its corresponding 2-D images, but the presented model just provides a possible morphological solution for the PN IRAS 21282. 
We should note that because of projection effects, it is possible that different 3-D models can lead to the images with a similar appearance. Only position-velocity (PV) diagrams obtained from spectroscopic observation can break this degeneracy. Further observations with integral field spectroscopy on large optical telescopes are needed to constraint our model.

\begin{table*}
	\caption{Properties of studied multipolar and quadrupolar planetary nebulae}
	\centering
	\begin{tabular}{ccccc}
		\hline\hline
PN G &  Name &  Spectral type$^{\sharp}$ & Binarity  \\
		\hline
002.6-03.4 & M 1-37 & [WC 11]$^a$ & No$^b$  \\
008.3-07.3 & NGC 6644 & [WEL]$^a$ & Possible$^c$  \\
009.6+14.8 & NGC 6309 & [WEL]$^a$ & --  \\
019.4-05.3 & M 1-61 & [WEL]$^a$ & --  \\
{021.8-00.4} & {M 3-28} & -- & {Possible$^d$}  \\
024.8-02.7 & M 2-46 & -- & --  \\
037.8-06.3 & NGC 6790 & [WN]$^a$ & -- \\
{043.0-03.0} & {M 4-14} & -- & {Possible$^d$}  \\
048.7+02.3 & K 3-24 & -- & {Possible$^d$} \\
064.6+48.2 & NGC 6058 & O9$^a$ & --  \\
068.8-00.0 & M 1-75 & -- & Possible$^e$  \\
074.5+02.1 & NGC 6881 & -- & {Possible$^f$} \\
084.7-08.0 & Kn 26 & -- & Possible$^g$  \\
089.0+00.3 & NGC 7026 & [WO 3]$^a$ & --  \\
089.8-05.1 & IC 5117 & [WR]$^a$ & --  \\
093.9-00.1 & IRAS 21282 & [WC 11]$^a$ & Possible$^{h,i}$ \\
234.8+02.4 & NGC 2440 & -- & -- &  \\
285.6-02.7 & Hen 2-47 & [WC 10-11]$^a$ & --  \\
300.7-02.0 & Hen 2-86 & [WC 4]$^a$ & --  \\
307.2-03.4 & NGC 5189 & [WO 1]$^a$ & Yes$^j$  \\
309.1-04.3 & NGC 5315 & [WO 4]$^a$ & Possible$^j$ \\
321.0+03.9 & Hen 2-113 & [WC 10]$^a$ & No$^j$  \\
332.9-09.9 & Hen 3-1333 & [WC 10]$^a$ & No$^j$  \\
355.9-04.2 & M 1-30 & [WEL]$^a$ & -- \\
359.3-00.9 & Hb 5 & -- & --  \\
		\hline
	\end{tabular}
%	\tablefoot{
%       \tablenotetext{$\sharp$}{[WEL]: weak emission-line stars.}
	\tablenotetext{a}{\citet{Weidmann11}}
        \tablenotetext{b}{\citet{Lutz10}}
        \tablenotetext{c}{\citet{Hsia10}}	
        \tablenotetext{d}{Manchado et al.(1996)}
        \tablenotetext{e}{Santander-Garc\'{i}a et al. (2010)}
        \tablenotetext{f}{Guerrero \& Manchado (1998)}
    \tablenotetext{g}{Guerrero et al. (2013)}
    \tablenotetext{h}{This study}
    \tablenotetext{i}{Cohen \& Jones (1987)}
    \tablenotetext{j}{\citet{Manick15}}  
    %    \tablenotetext{i}{Phillips et al. (2003).}
%    \tablenotetext{j}{Hsia et al. (2014).}
%    \tablenotetext{k}{Stanghellini et al. (2002).}
%   \tablenotetext{l}{Assuming largest expansion velocity of the lobes to be equal to wind value of the object.}    
%    }
	\label{tab7}
\end{table*}

\section{Comparison with other multipolar planetary nebulae}\label{s6}

When stars evolve from AGB to PN stages, some develop hydrogen-deficient atmospheres above their C-O electron-degenerate cores. But the exact mechanism removing hydrogen is still an open issue. The spectra of these objects always mimics the massive Wolf-Rayet (WR) star exhibiting strong stellar wind with mass-loss rate up to about 10$^{-6}$ M$_\sun$~yr$^{-1}$ \citep{Crowther08}. 
Based on the strengths of emission lines, the visual spectral classification of [WR]-type central star of PN (CSPN) can be classified as [WC] (carbon-rich), [WN] (nitrogen-rich), and [WO] (oxygen-rich) subtypes \citep{Crowther08}. The hydrogen atmosphere could be removed at (i) at the end of the AGB stage (AGB final thermal pulse; Herwig 2001); (ii) at the post-AGB stage (late thermal pulse; Bl\"{o}cker 2001); or (iii) at white dwarf cooling track  (very late thermal pulse; Lawlor \& MacDonald 2002). 
An hypothesis for the origin of [WR]-type central stars is by binary/planet interaction \citep{De02, De08}, where a companion or a massive planet is engulfed by an AGB star. 

IRAS 21282 has been identified as a multipolar PN with the cool [WC 11] nucleus (see Section~\ref{spectra}). It would be interesting to investigate whether the multipolar structures are related to the [WR] nature of the central star. 
For that purpose, we collected a sample of twenty-three known multipolar PNs that has been studied in previous literatures \citep{Clark13, Dan15, Guerrero98, Guerrero13, Guillen13, Hsia10, Hsia14, Lopez98, Lopez12, Manchado96, 
Sabin12, Sahai00, Vazquez08}. 
The adopted physical parameters of these sources are listed in Table \ref{tab7}. The columns 1 and 2 give the PNG numbers and object names. The spectral types of central sources and binarity of them are given in columns 3 and 4. %Column 5 gives the effective temperatures of central stars ($T_{eff}$). Largest expansion velocity of the lobes ($V_{lobe}$) and those obtained from the references are listed in columns 6 and 7, respectively. 

From Table \ref{tab7}, we find that that a significant fraction ($\sim$ 44$\%$) of multipolar PNs in our sample shows [WR]-type nucleus and 4 objects (M 1-30, M 1-61, NGC 6309, and NGC 6644) have [WEL]-type central stars.
Possible explanations to such correlations include  multiple lobes produced by strong turbulence blowing from [WC]-type central stars (e.g. Acker et al. 2002; Gesicki et al. 2003) or atmospheric instabilities in the shells 
during the AGB stages \citep{steffen13}. 

The possible presence of  binary nuclei detected in multipolar PNs (Table \ref{tab7}) has led to suggestion that precessions of mass-losing star's rotation axis and jets with time-dependent ejection from the binary or multiple sub-stellar companions in the nuclei could need to the formation of multipolar PNs \citep{Garcia97, Velazquez12}. 
It would be interesting to investigate further whether the binary nature of the central star of IRAS 21282 plays any role in the observed multipolar structure of this object.
%It is possible the present sample of multipolar PNs with central binaries represents the tip of the iceberg, further measurement is needed to confirm the scenario.  

%A comparison of the correlation between central-star temperatures ($T_{eff}$) versus largest lobe expansion velocities ($V_{lobe}$) of these nebulae is given in Figure~\ref{vlobe}. The slope of 0.72 and correlation coefficient of 0.75 for this plot in the linear space are derived. From Figure~\ref{vlobe}, no obvious difference is found in the trend of these [WR]-type PNs compared to that of the other PNs. A clear tendency shows that largest expansion velocity of the lobes strongly increases with $T_{eff}$, suggesting that nebular acceleration is closely related to evolutionary age, as traced by the temperature of central star. This is consistent with the previous results suggested by \citet{Medina06}.

%\begin{figure}
%        \begin{center}
%                \includegraphics[width=0.52\textwidth]{f8.eps}
%        \end{center}
%        \caption{Central-star temperatures (T$_{eff}$) vs. lobe expansion velocities (V$_{lobe}$) for studied 15 multipolar PNs. The dotted line is a linear fit to the data. The fitting parameters are shown on the upper left corner of the figure.   The open diamonds, filled triangles and circles are multipolar non-WR CSPNs, WR CSPNs, and WEL CSPNs, respectively.  The derived correlation coefficient for this plot is 0.75.
%        \label{vlobe}}
%\end{figure}

\section{Conclusions}

From {\it HST} high-angular-resolution optical and near-infrared observations, we have identified at least three pairs of bipolar lobes outside the main elliptical shell in the PN IRAS 21282.  In addition, we find an infrared torus which is approximately perpendicular to the elliptical shell.  A cylinderical shape structure found in the optical image could represent a fourth pair of bipolar lobes with a symmetry axis perpendicular to the dust torus.

The optical spectrum of the central star exhibits TiO absorption bands. Different emission lines show different RVs, which can be roughly divided into two groups. These results suggest the presence of a cool companion.  A number of circumstellar DIBs are seen in the optical spectrum.  It would be interesting to investigate whether the presence of the DIBs are related to the circumstellar UIE bands.

In order to illustrate the 3-D intrinsic structures of IRAS 21282, we have constructed a model that allows the visualization of the PN viewed from different orientations. The simulated rotated images of this object show that IRAS 21282 may have similar intrinsic structures as other multipolar PNs, suggesting that each multipolar PN is not unique and multipolar structures are common among PNs.

A significant fraction ($\sim$ 44$\%$) of multipolar PNs in our sample is found to have [WR]-type central stars.  
%, indicating that the complex multiple lobes may be produced by strong turbulence blowing from central [WC]-type nuclei, atmospheric instabilities in the AGB shells, and binary/planet interactions. 
It would be interesting to explore whether the multi-nebular properties of these nebulae are related to the spectral properties of the central sources.
%For all types of multipolar PN, the largest expansion velocity of the lobes is increasing with the central-star temperature.

IRAS 21282 is another example of the growing list of multipolar PNs found as the result of higher sensitivity and dynamical range imaging. It is possible that the multipolar phenomenon is much more common than previously believed and multipolar nature of PNs may be a norm rather than an exception.

\acknowledgments
We thank an anonymous referee for his/her helpful comments.  
Part of the data presented in this paper were obtained from the Multi-mission Archive at the Space Telescope Science Institute (MAST). STScI is operated by the Association of Universities for Research in Astronomy, Inc., under NASA contract NAS5-26555. Support for MAST for non-HST data is provided by the NASA Office of Space Science via grant NAG5- 7584 and by other grants and contracts. 
We acknowledge the support of the staff of the Xinglong 2.16m telescope. This work was partially supported by the Open Project Program of the Key Laboratory of Optical Astronomy, National Astronomical Observatories, Chinese Academy of Sciences. 
The Laboratory for Space Research was established by a grant from the University Development Fund of the University of Hong Kong. 
Financial support for this work is supported by the grants from Science and Technology Development Fund of Macau (project codes: 119/2017/A3 and 061/2017/A2) %HKRGC (HKU 7027/11P and HKU 7062/13P), 
and a grant to SK from the Natural and Engineering Research Council of Canada.

%% References
%% Please cite all reference entries in the article text using \cite or
%% equivalent command. 

%%%  Using BibTeX  (Name-Year style)
%
% \bibliographystyle{spr-mp-nameyear-cnd}  %% BibTeX style
% \bibliography{<bib data>}                %% BibTeX data

\begin{thebibliography}{}

\bibitem[Acker et al.(2002)]{Acker02}
Acker, A., Kesicki, K., Grosdidier, Y., \& Durand, S. 2002, \aap, 384, 628

\bibitem[Bailer-Jones et al.(2018)]{Bailer18}
Bailer-Jones, C. A. L., Rybizki, J., Fouesneau, M., Mantelet, G., \& Andrae, R., 2018, \aj, 156, 58

\bibitem[Balick(1987)]{Balick87}
Balick, B. 1987, \aj, 84, 671 

\bibitem[Bl\"{o}cker(2001)]{Blocker01}
Bl\"{o}cker, T. 2001, \apss, 275, 1

\bibitem[Casassus et al.(2001)]{cas01} 
Casassus, S., Roche, P.~F., Aitken, D.~K., \& Smith, C.~H.\ 2001, \mnras, 320, 424 

\bibitem[Castro-Carrizo et al.(2010)]{Castro10}
Castro-Carrizo, A., Quintana-Lacaci, G., Neri, R., et al. 2010, \aap, 523, A59

%\bibitem[Castro-Carrizo et al.(2012)]{Castro12}
%Castro-Carrizo, A., Neri, R., \& Bujarrabal, V., et al. 2012, \aap, 545, 1

\bibitem[Cheng(2005)]{Cheng05}
Cheng, J. 2005, High-Resolution Imaging of Objects in the Post-AGB to Planetary Nebula Transition, 
%\url{http://reu.physics.ucla.edu/common/papers/2005/judy_cheng.pdf}

\bibitem[Chong et al.(2012)]{chong2012}
Chong, S.-N., Kwok, S., Imai, H., Tafoya, D., \& Chibueze, J. 2012, ApJ, 760, 115

\bibitem[Clark et al.(2013)]{Clark13}
Clark, D. M., L\'{o}pez, J. A., Steffen, W., et al. 2013, \aj, 145, 57 

\bibitem[Cohen, Tielens, \& Allamandola(1985)]{Cohen85}
Cohen, M., Tielens, A. G. G. M. \& Allamandola, L. J. 1985, \apjl, 299, 93

\bibitem[Cohen \& Jones(1987)]{Cohen87}
Cohen, M., \& Jones, B. F. 1987, \apjl, 321, 151

\bibitem[Crowther(2008)]{Crowther08}
Crowther, P. A., 2008, in Hydrogen-Deficient Stars, ed. K. Werner, \& T. Rauch, ASP Conf. Ser., 391, 83

\bibitem[Cutri et al.(2003)]{Cutri03}
Cutri, R. M., et al. 2003, VizieR Online Data Catalog: II/246

\bibitem[Cutri et al.(2012)]{Cutri12}
Cutri, R. M., Wright, E. L., Conrow, T. A. O. et al. 2012, VizieR Online Data Catalog: II/311

%\bibitem[Crowther et al.(1998)]{Crowther98}
%Crowther, P. A., De Marco, O., \& Barlow, M. J. 1998, \mnras, 296, 367

\bibitem[Danehkar \& Parker(2015)]{Dan15}
Danehkar, A., \& Parker, Q. A. 2015, \mnras, 449, L56

\bibitem[Davis et al.(2005)]{Davis05}
Davis, C. J., Smith, M. D., Gledhill, T. M., et al. 2005, \mnras, 360, 104

\bibitem[De Marco(2008)]{De08}
De Marco, O., 2008, in Hydrogen-Deficient Stars, ed. K. Werner, \& T. Rauch, ASP Conf. Ser., 391, 209  

\bibitem[De Marco \& Soker(2002)]{De02}
De Marco, O., \& Soker, N. 2002, \pasp, 114, 602 

\bibitem[Egan et al.(2003)]{Egan03}
Egan, M. P., Price, S. D., Kraemer, K. E., et al. 2003, The Midcourse Space Experiment Point Source Catalog v2.3, Air Research Laboratory 
Technical Report AFRL-VS-TR- 2003-1589


%\bibitem[Fong et al.(2001)]{Fong01}
%Fong, D., Meixner, M., Castro-Carrizo, A., et al. 2001, \aap, 367, 652 


\bibitem[Garc\'{i}a-Segura(1997)]{Garcia97}
Garc\'{i}a-Segura, G. 1997, \apjl, 489, L189


\bibitem[Garc\'{i}a-Segura(2010)]{Garcia10}
Garc\'{i}a-Segura, G. 2010, \aap, 520, L5

\bibitem[Garc\'{i}a-Hern\'{a}ndez \& D\'{i}az-Luis(2013)]{Garcia13}
Garc\'{i}a-Hern\'{a}ndez, D. A. \& D\'{i}az-Luis, J. J. 2013, \aap, 550, 6

\bibitem[Gesicki et al.(2003)]{Gesicki03}
Gesicki, K., Acker, A., \& Zijlstra, A. A. 2003, \aap, 400, 957

\bibitem[Guerrero \& Manchado(1998)]{Guerrero98}
Guerrero, M. A., \& Manchado A. 1998, \apj, 508, 262 

\bibitem[Guerrero et al.(2013)]{Guerrero13}
Guerrero, M. A., Miranda, L. F., Ramos-Larios, G., \& V\'{a}zquez, R. 2013, \aap, 551, 53 

\bibitem[Guill\'{e}n et al.(2013)]{Guillen13}
Guill\'{e}n, P. F., V\'{a}zquez, R., Miranda, L. F., et al. 2013, \mnras, 432, 2676

\bibitem[Herwig (2001)]{Herwig01}
Herwig, F., 2001, \apss, 275, 15

%\bibitem[Hillwig et al.(2015)]{Hillwig15}
%Hillwig, T. C., Frew, D. J., Louie, M., et al. 2015, \aj, 150, 30

\bibitem[Hora, Latter, \& Deutsch(1999)]{Hora99}
Hora, J. L., Latter, W. B., \& Deutsch, L. K. 1999, \apjs, 124, 195

\bibitem[Hrivnak, Geballe \& Kwok(2007)]{Hrivnak07}
Hrivnak, B. J., Geballe, T. R., \& Kwok, S. 2007, \apj, 662, 1059

\bibitem[Hsia et al.(2010)]{Hsia10}
Hsia, C.-H., Kwok, S., Zhang, Y., et al. 2010, \apj, 725, 173

\bibitem[Hsia et al.(2014)]{Hsia14}
Hsia, C.-H., Chau, W., Zhang, Y., \& Kwok, S. 2014, \apj, 787, 25

\bibitem[Hsia et al.(2016)]{Hsia16}
Hsia, C.-H., Sadjadi, S., Zhang, Y., \& Kwok, S. 2016, \apj, 832, 213

\bibitem[Ishihara et al.(2010)]{Ishihara10}
Ishihara, D., et al. 2010, \aap, 514, A1

\bibitem[Ivezic, Nenkova, \& Elitzur(1999)]{Ivezic99}
Ivezic , Z., Nenkova, M., \& Elitzur M. 1999, DUSTY user manual, University of Kentucky internal report

\bibitem[Jourdain de Muizon et al.(1986)]{Jourdain86}
Jourdain de Muizon, M., Geballe, T. R., D'Hendecourt, L. B., et al. 1986, \apjl, 306, 105

\bibitem[Jourdain de Muizon, D'Hendecourt, \& Geballe(1990)]{Jourdain90}
Jourdain de Muizon, M., D'Hendecourt, L. B., \& Gabelle, T. R. 1990, \aap, 227, 526

\bibitem[Kimeswenger \& Barr{\'{\i}}a(2018)]{kim18} 
Kimeswenger, S., \& Barr{\'{\i}}a, D.\ 2018, \aap, 616, L2



\bibitem[Knapp et al.(1995)]{Knapp95}
Knapp, G. R., Bowers, P. F., Young, K., \& Phillips, T. G. 1995, \apj, 455, 293

\bibitem[Kwok, Hrivnak, \& Langill(1993)]{Kwok93}
Kwok, S., Hrivnak, B. J., \& Langill, P. P. 1993, \apj, 408, 586

\bibitem[Kwok, Volk, \& Hrivnak(1999)]{Kwok99}
Kwok, S., Volk, K., \& Hrivnak, B. J. 1999, \aap, 350, 35

\bibitem[Kwok \& Su(2005)]{Kwok05}
Kwok, S., \& Su, K. Y. L. 2005, \apjl, 635, 52

\bibitem[Kwok(2007)]{Kwok07}
Kwok, S. 2007, Physics and Chemistry of the Interstellar Medium (Sausalito, CA:Univ. Science Books)

\bibitem[Kwok(2010)]{Kwok10a}
Kwok, S. 2010, \pasa, 27, 174

\bibitem[Kwok et al.(2010)]{Kwok10b}
Kwok, S., Chong, S.-N., Hsia, C.-H., et al. 2010, \apj, 708, 93

%\bibitem[Kwok(2011)]{Kwok11}
%Kwok, S. 2011, in IAU Symp. 280, The Molecular Universe, ed. J. Cernicharo \& R. Bachiller (Cambridge: Cambridge Univ. Press)

\bibitem[Lawlor \& MacDonald(2002)]{Lawlor02}
Lawlor, T. M., \& MacDonald, J. 2002, in Exotic Stars as Challenges to Evolution, ed. C. A. Tout, \& W. Van Hamme, ASP Conf. 
Ser., 279, 193

\bibitem[Le Bertre \& Lequeux(1992)]{Le92}
Le Bertre, T., \& Lequeux, J. 1992, \aap, 255, 288

\bibitem[Leuenhagen \& Hamann(1998)]{Leuenhagen98}
Leuenhagen, U., \& Hamann, W.-R. 1998, \aap, 330, 265

%\bibitem[Likkel et al.(1988)]{Likkel88}
%Likkel, L., Morris, M., Forveille, T., et al. 1988, \aap, 198, 1
\bibitem[Likkel(1988)]{lik88} 
Likkel, L., Forveille, T., Omont, A., \& Morris, M.\ 1988, \aap, 198, L1 

\bibitem[Likkel et al.(1994)]{Likkel94}
Likkel, L., Morris, M., Kastner, J. H., et al. 1994, \aap, 282, 190

%\bibitem[Liu et al.(2001)]{Liu01}
%Liu, X.-W., Barlow, M. J., Cohen, M., et al. 2001, \mnras, 323, 343

\bibitem[L\'{o}pez et al.(1998)]{Lopez98}
L\'{o}pez, J. A., Meaburn, J., Bryce, M., \& Holloway, A. J. 1998, \apj, 493, 803

\bibitem[L\'{o}pez et al.(2012)]{Lopez12}
L\'{o}pez, J. A., Garc\'{i}a-D\'{i}az, M. T., Steffen, W., Riesgo, H. \& Richer, M. G. 2012, \apj, 750, L131

\bibitem[Lutz et al.(2010)]{Lutz10}
Lutz, J., Fraser, O., McKeever, J., \& Tugaga, D. 2010, \pasp, 122, 524

\bibitem[Manchado, Stanghellini, \& Guerrero(1996)]{Manchado96}
Manchado, A., Stanghellini, L., \& Guerrero, M. A. 1996, \apj, 466, L95

\bibitem[Manchado et al.(2011)]{Manchado11}
Manchado, A., Garc\'{i}a-Hern\'{a}ndez, D. A., Villaver, E., et al. 2011, in ASP Conf. Ser. 445, Why Galaxies Care About AGB Stars II, ed. F. Kerschbaum, T. Lebzelter, \& B. Wing (San Francisco, CA:ASP). 161  

\bibitem[Manick et al.(2015)]{Manick15}
Manick, R., Miszalski, B., \& McBride, V. 2015, \mnras, 448, 1789

\bibitem[Mastrodemos \& Morris(1998)]{Mastrodemos98}
Mastrodemos, N., \& Morris, M. 1998, \apj, 497, 303 


%\bibitem[Medina et al.(2006)]{Medina06}
%Medina, S., Pe\~{n}a, M., Morisset, C., \& Stasi\'{n}ska, G. 2006, RMxAA, 42, 53
 
\bibitem[Meixner et al.(1993)]{Meixner93}
Meixner, M., Skinner, C. J., Temi, P., et al. 1993, \apj, 411, 266


\bibitem[Moshir et al.(1992)]{Moshir92}
Moshir, M., Kopman, G., \& Conrow, T. A. O. (ed.) 1992, IRAS Faint Source Survey, Explanatory Supplement Version 2

\bibitem[Nagata et al.(1988)]{Nagata88}
Nagata, T., Tokunaga, A. T., \& Sellgren, K., 1988, \apj, 326, 157 

%\bibitem[Phillips et al.(2003)]{Phillips03}
%Phillips, J. P., 2003, \mnras, 344, 501

%\bibitem[Rechy-Garc\'{i}a et al.(2017)]{Rechy17}
%Rechy-Garc\'{i}a, J. S., Vel\'{a}zquez, P. F., Pe\~{n}a, M, \& Raga, A. C. 2017, \mnras, 464, 2318 

\bibitem[Rubio et al.(2015)]{Rubio15}
Rubio, G., V\'{a}zquez, R., Ramos-Larios, G., et al. 2015, \mnras, 446, 1931 

\bibitem[Sabin et al.(2012)]{Sabin12}
Sabin, L., V\'{a}zquez, R., L\'{o}pez, J. A., et al. 2012, RMxAA, 48, 165

\bibitem[Sahai(2000)]{Sahai00}
Sahai, R. 2000, \apjl, 537, 43

\bibitem[Sahai, Morris, \& Villar(2011)]{Sahai11}
Sahai, R., Morris, M. R., \& Villar, G. G. 2011, \aj, 141, 134

\bibitem[Santander-Garc\'{i}a et al.(2010)]{Santander10}
Santander-Garc\'{i}a, M., Rodr\'{i}guez-Gil, P., Hernandez, O., et al. 2010, \aap, 519, A54 

%\bibitem[Sch\"{o}nberner(1981)]{Schonberner81}
%Sch\"{o}nberner, D. 1981, \aap, 103, 119

\bibitem[Shibata et al.(1989)]{shi89} 
Shibata, K.~M., Tamura, S., Deguchi, S., et al.\ 1989, \apjl, 345, L55

%\bibitem[Stanghellini et al.(2002)]{Stanghellini02}
%Stanghellini, L., Villaver, E., Manchado, A., \& Guerrero, M. A. 2002, \apj, 576, 285

\bibitem[Stanghellini et al.(2017)]{sta17} 
Stanghellini, L., Bucciarelli, B., Lattanzi, M.~G., \& Morbidelli, R.\ 2017, \na, 57, 6


\bibitem[Steffen et al.(2011)]{steffen11}
Steffen, W., Koning, N., Wenger, S., Morisset, C., \& Magnor, M. 2011, IEEE Trans. Vis. Comput. Graphics, 17, 454

\bibitem[Steffen et al.(2013)]{steffen13}
Steffen, W., Koning, N., Esquivel, A., et al. 2013, \mnras, 436, 470

\bibitem[Su et al.(2004)]{Su04}
Su, K. Y. L., Kelly, D. M., Latter, W. B., et al. 2004, \apjs, 154, 302

%\bibitem[Taylor et al.(1987)]{Taylor87}
%Taylor, A. R., Pottasch, S. R., \& Zhang, C. Y. 1987, \aap, 171, 178 

\bibitem[van der Hucht et al.(1981)]{van81}
van der Hucht, K. A., Conti, P. S., Lundstrom, I., et al. 1981, \ssr, 28, 227 

\bibitem[V\'{a}zquez et al.(2008)]{Vazquez08}
V\'{a}zquez, R., Miranda, L. F., Olgu\'{i}n, L., et al. 2008, \aap, 481, 107

\bibitem[Vel\'{a}zquez et al.(2012)]{Velazquez12}
Vel\'{a}zquez, P. F., Raga, A. C., Riera, A., et al. 2012, \mnras, 419, 3529

\bibitem[Wright et al.(2010)]{Wright10}
Wright, E. L., Eisenhardt, P. R. M., Mainzer, A. K., et al. 2010, \aj, 140, 1868

\bibitem[Weidmann \& Gamen(2011)]{Weidmann11}
Weidmann, W. A., \& Gamen, R. 2011, \aap, 526, A6


\bibitem[Zhang \& Kwok(1991)]{Zhang91}
Zhang, C. Y., \& Kwok, S. 1991, \aap, 250, 179



\end{thebibliography}

%% Non-BibTeX  (Name-Year style)
%
% \begin{thebibliography}{}
% \bibitem[\protect\citeauthoryear{<author>}{<year>]{ref:?}
%    <ref. entry>
% \bibitem[\protect\citeauthoryear{<author>}{<year>]{ref:?}
%    <ref. entry>
% \end{thebibliography}

\end{document}